\documentstyle[aps,prl,amssymb,rotating,epsfig,color,graphics,eepic,twocolumn]{revtex}
\newcommand{\be}{\begin{equation}}
\newcommand{\ee}{\end{equation}}
\newcommand{\bea}{\begin{eqnarray}}
\newcommand{\eea}{\end{eqnarray}}

\newcommand{\cl}{{\cal{L}}}

\begin{document}
\twocolumn[\hsize\textwidth\columnwidth\hsize\csname@twocolumnfalse\endcsname
\title{Master Equation Approach to Molecular Motors}
\author{G Lattanzi$^{*1}$ \and A Maritan$^{1,2}$}
\address{$^*$ Corresponding author: lattanzi@sissa.it \\
$^1$ International School for Advanced Studies and 
Istituto Nazionale di Fisica della Materia, \\
Via Beirut 2-4, 34014 Trieste, Italy and \\ 
$^2$ The Abdus Salam International Center for Theoretical Physics, \\
Strada Costiera 11, 34100 Trieste, Italy.}
\date{Editorially approved for publication on Physical Review E, Aug 22, 2001.} 

\bibliographystyle{unsrt}

\maketitle

\begin{abstract}

A master equation approach to molecular motors allows to describe a
mechano--chemical cyclic system where chemical and translational degrees of
freedom are treated on an equal footing. A generalized detailed balance condition 
in the out of equilibrium regime is shown to be compatible with the
Fokker--Planck equation in the continuum limit. The Onsager reciprocity 
relations hold for stationary states close to equilibrium, provided the 
generalized detailed balance condition is satisfied. 
Semi--phenomenological considerations in the case of motor proteins lead to a 
discrete kinetics model, for which interesting observable quantities can be 
directly calculated and compared with experimental data.

\vspace{5pt}
\vspace{5pt}

\end{abstract}
]

Current models used to describe the properties of molecular motors and the 
energy transduction process fall under two distinct categories: continuous 
models~\cite{mag93,jap97,par99} and discrete models~\cite{fish99,lip00}. Both 
represent a coarse grained description of a very complicated physico--chemical 
system and the use of one or the other depends on the quantities one is 
interested in. 
For example, continuous models are very useful to investigate the role of an 
external force on chemical kinetics~\cite{lat01}, since the external 
force is inserted into the Fokker--Planck (FP) equations without any ambiguity.
This is no longer true 
for discrete models, when one has to resort to some ad hoc principle or a 
priori reasoning to insert force in transition rates~\cite{fish99}. 
Nonetheless discrete models present the important advantage of being 
analytically solvable, as it happens, for instance, in jump 
processes~\cite{der83}.
On the other hand an analytical solution is quite difficult to obtain in the 
general case of continuous models, and one has to resort to
complex numerical integrations. 
In this paper we introduce a discrete model, similar to the 
ones proposed in~\cite{fish99} and~\cite{lip00}, but with the following
constraint: 
if $a$ is the lattice distance between subsequent spatial positions of the 
system, the continuous model should be obtained as a limit of the discrete one 
for $a\to 0$.  
The connection between a kinetic theory involving activated transitions over
potential energy barriers and a diffusion theory approach based on a
FP equation dates back to Kramers~\cite{kram40}, see~\cite{han90}
for a recent review. 
In~\cite{ast96}, the idea was applied to models for protein
motors, but the force dependence was left in equal apportionments over backward
and forward transition rates and the experimental results on the force
dependence of the apparent Michaelis constant~\cite{viss99} were not available.
In~\cite{kel00}, a general theory for motor proteins was presented. This
theory was developed in a two dimensional manifold and complex integrations over
state variables were used to calculate force dependent transition rates over
potential barriers in a discrete model. In the present paper we do not use
complex integrations; instead, we identify in the generalized detailed balance
the condition for a discrete model to be compatible with a continuous one. Few
parameters are needed to capture the overall shape of the potential energy
surface which is no longer needed in full detail.
Detailed balance was used in~\cite{els96} to calculate transition rates for
particles diffusing over potential barriers. The obtained discrete kinetics 
model led to a fast and reliable numerical procedure to calculate mean velocity
of correlation ratchets.
A model similar to ours was also developed in~\cite{sok97} in the context of 
thermal ratchets, even if no connection with the actual chemistry of motor 
proteins was done. Moreover, motor proteins are isothermal, and therefore 
are better described by correlation ratchets.
In~\cite{wang98}, a master equation approach was used to investigate the force
generation in RNA polymerase, which can be considered a motor protein, even if
it differs from kinesin, myosin and dynein both in structure and function.
In section~\ref{sec:gf} we outline the general framework, which may be useful
not only for modeling molecular motors, but also any mechano--chemical 
cyclic system. In the general formulation of our model, the chemical reaction 
coordinate is treated on an equal footing as the spatial one. Onsager
reciprocity relations will be shown to hold in the most general case in the 
stationary periodic close to equilibrium state, provided detailed balance is 
verified. In section~\ref{sec:cl} we specify our model to the context of 
molecular motors. We show that our model may be regarded as the discrete 
analogue of the continuous one proposed in~\cite{jap97,par99}, leading to a 
clear interpretation of the generalized forces and currents introduced in 
section~\ref{sec:gf}. 
In section~\ref{sec:def} a discrete chemical kinetics model with a generalized
detailed balance condition is defined for the case of motor proteins.
Semi--phenomenological considerations help to decide the apportionments of
generalized forces over forward and backward transitions. In
section~\ref{sec:mpaes} two example models are studied and their predictions
compared with experimental data.

\section{General framework}
\label{sec:gf}

Our model will describe the time evolution of a thermodynamic 
out--of--equilibrium system in a complex phase space. 
The state of the system (a motor protein, an ion pump or whatsoever) is
determined by the thermodynamic parameters $x_\alpha (\alpha = 1, \ldots , d)$.
One of these parameters may represent the position of the protein center of 
mass, another a chemical reaction coordinate (or a variable indicating the 
conformation of the protein) and so on. In general the number of these 
thermodynamic parameters (and hence the dimensionality of the system under 
study) is sufficient to identify the state of the system by a direct 
experimental measure.
Therefore we will assume the state of the system to be described by a 
$d-$dimensional vector $X$ subject to a time evolution in a 
$d-$dimensional discrete phase space, which can be mapped on ${\cal{Z}}^d$ (in
general the lattice spacing in each direction will be different).

The probability of being in a particular state--$X$ at time $t$ is written 
as $P_X(t)$.
Since the system must be in one of the $X$ states, the normalization condition 
follows:

\be
\sum\nolimits_X P_X(t) = 1 \quad \forall \: t \: \label{eq:norm}
\ee

$W_{XY}$ is defined as the transition probability per unit time from state $Y$ 
to state $X$ and it is assumed to be time independent. 

Another hypothesis is the full periodicity along any direction $\alpha$. 
This is usually assumed for all models of motor 
proteins, (see~\cite{ast96,kel00}). The periodicity $N_\alpha $  
depends on $\alpha $, but we assume $N_\alpha \ge 1 $, since 
it is always possible to reduce the steps until this constraint is satisfied.  
In other words we assume that the state described by the parameters $(l_1 N_1+x_1, 
\ldots , l_d N_d+x_d)$ with $L \equiv \{l_1, \ldots l_d \} \in {\cal{Z}}^d$ is 
equivalent to the state described by the parameters $(x_1,\ldots ,x_d)$.

We introduce the variable:

\be
\chi_{_X} = 
\left\{
\begin{array}{ll}
1 & \mbox{for } 1\le x_\alpha < N_\alpha \quad \forall \, \alpha \, \in \{ 1,
\ldots , d \} \\
0 & \mbox{otherwise}
\end{array}
\right.
,
\label{eq:defchi}
\ee

\noindent which is an indicator of the period in which the system is moving and 
will be useful for subsequent calculations. 

The time evolution of the system is simply given by the master equation:

\be
\dot{P_X} = \sum\nolimits_Y \left( W_{XY} P_Y - W_{YX}P_X \right)\equiv \sum\nolimits_Y 
L_{XY} P_Y ,
\label{eq:master}
\ee

\noindent where we have defined:

\be
L_{XY} = W_{XY} - \delta_{XY} \sum\nolimits_Z W_{ZX}.
\label{eq:deflxy}
\ee

From this definition, it follows:

\be
\sum\nolimits_X L_{XY} = 0 \Longrightarrow \sum\nolimits_X \dot{P_X} = 0 ,\label{eq:sumx}
\ee

\noindent which is consistent with the normalization condition, eq.~(\ref{eq:norm}). 
Since the system is assumed to be periodic,

\be
W_{X+LN,Y+LN} = W_{XY} \quad \forall L \in {\cal{Z}}^d,
\ee

\noindent where $LN \equiv (l_1N_1,\ldots,l_dN_d)$. It is simple to show that the same 
rule holds also for the matrix $L_{XY}$, defined in eq.~(\ref{eq:deflxy}).

We introduce a time independent variable $q_X$, 
depending explicitly on the $X$ coordinate, i.e. on the state 
of the system. We define the current $J_q$ conjugated to the variable $q_X$:

\be
J_q  = \frac{d\left< q \right>}{dt},
\ee

\noindent where $\left< q \right>$ is the average: $\left< q \right> = 
\sum\nolimits_X q_X P_X$. By applying eqs.~(\ref{eq:master}) and~(\ref{eq:sumx})
it easily follows:

\be
J_q  = \sum\nolimits_{XY}\left( q_X - q_Y \right) L_{XY}P_Y.
\ee

We introduce probabilities and transition rates over all periods, 
following some of the formalism of the $d=1$ case studied in \cite{der83}:

\bea
R_X & \equiv & \sum\nolimits_L P_{X+LN} \label{eq:defr}\\
\cl_{XY} & \equiv & \sum\nolimits_L L_{X,Y+LN} \label{eq:defl}.
\eea

By definition, $R_X$ is a periodic quantity. It is easy to show 
that also $\cl_{XY}$ is periodic in both arguments $X$ and $Y$. 
At variance of $L_{XY}$, $\cl_{XY}$ is a finite matrix; also 
$R_X$ is a finite vector, whereas $P_X$ is not.
From  the time evolution of $P_X$, we can easily obtain the time evolution of 
$R_X$:

\be
\dot{R_X} = \sum\nolimits_Y L_{XY} R_Y.
\label{eq:te1}
\ee

For any variable $f_Y$ and using eq.~(\ref{eq:defchi}) the following property 
holds:

\be
\sum\nolimits_X f_X = \sum\nolimits_X \chi_{_X} \sum\nolimits_L f_{X+LN}.
\label{eq:prop}
\ee

Applying property~(\ref{eq:prop}) to eq.~(\ref{eq:te1}) and the periodicity of
$R$, we obtain:

\be
\dot{R_X} = \sum\nolimits'_Y \cl_{XY} R_Y,
\label{eq:te}
\ee

\noindent where, by definition, $\sum\nolimits'_Y = \sum\nolimits_Y \chi_{_Y} $, 
i.e. a primed sum is restricted only to one period along any axis.
This is a master equation for a system with a finite number $\left(
\prod_{\alpha=1}^{d} N_\alpha \right) $ of states. The  
$\cl_{XY}$ matrix is finite and has the following properties:

\bea
\cl_{XY} & \ge & 0 \; \; \mbox{for} \, X \not= Y \\
\sum\nolimits_{X}' \cl_{XY} & = & 0 \; \; \forall \, Y \, \in \, {\cal{Z}}^d.
\eea

It is easy to show, by applying eqs. (\ref{eq:prop}) and (\ref{eq:sumx}), that:

\be
\sum\nolimits'_X R_X = \sum\nolimits_X P_X = 1.
\ee

Therefore, there exists a stationary solution $\widehat{R}_X$ of eq.(\ref{eq:te})
and, under  general hypotheses (always satisfied in the examples treated in 
the next sections), it is unique~\cite{kamp81}. It is also periodic, by 
definition. 

If $q_X = x_\alpha$ for any value of $\alpha $, then $q_X-q_Y=q_{X+LN}-
q_{Y+LN}$ is also periodic, and we can rewrite the current as:

\be
J_q = \sum\nolimits_{XY} \chi_{_Y} \left( q_X -q_Y \right) L_{XY} R_Y \quad , \quad
q_X =x_\alpha.
\label{eq:current}
\ee

A subsequent application of property~(\ref{eq:prop}) to eq.~(\ref{eq:current}) 
gives:

\be
J_q = \sum\nolimits_{XY} \chi_{_X} \left( q_X -q_Y \right) L_{XY} R_Y \quad , \quad
q_X =x_\alpha
\ee

\noindent and, after some manipulation, we obtain the following expression:

\be
J_q = \sum\nolimits_{XY} \frac{\chi_{_X} + \chi_{_Y}}{4} \left( q_X -q_Y
\right) \left( L_{XY} R_Y - L_{YX} R_X \right),
\label{eq:symj}
\ee

where the argument in the sum is evidently symmetric under a 
change $X \leftrightarrow Y$.

If the detailed balance condition for the periodic stationary state,
$R_X=\widehat{R}_X$, holds:

\be
W_{XY} \widehat{R}_Y = W_{YX} \widehat{R}_X,
\label{eq:db}
\ee

\noindent which is equivalent to the following:

\be
L_{XY} \widehat{R}_Y = L_{YX} \widehat{R}_X,
\label{eq:debal}
\ee

\noindent then from eq.~(\ref{eq:symj}) the net stationary current is 
zero\footnote{Notice that if eq.~(\ref{eq:debal}) holds, then also $\cl_{XY} 
\widehat{R}_Y = \cl_{YX} \widehat{R}_X $ holds, but the converse is not 
guaranteed to be true.}. A net flow, i.e. a nonzero 
stationary current, may occur only if the detailed balance condition, 
eq.~(\ref{eq:db}), is violated. This can be done in several ways: in the 
continuous model proposed in ref.~\cite{par99}, for instance, detailed balance 
holds separately for each chemical reaction, introducing the chemical potential 
$\Delta \mu$.
In our model we introduce a set of generalized forces, able to drive the system
out of equilibrium, so that a finite stationary current may occur.
Each generalized force, $f_\alpha$ is coupled to one generalized 
coordinate $x_\alpha$. Both the transition matrix $L_{XY}$ and the stationary
solution will depend explicitly on the force vector $F$, so that:

\be
\sum\nolimits'_Y \cl_{XY}(F) \widehat{R}_Y (F)= 0
\label{eq:st}
\ee 
 
Of course, at equilibrium $F=0$ and the stationary currents are all
identically zero. Our assumption is that condition~(\ref{eq:db}) is replaced by 
a generalized detailed balance condition (in this section and in
appendix~\ref{app:ons}, the factor $\beta=1/k_BT$ is absorbed in the definition
of $F$):

\be
L_{XY}(F) \widehat{R}_Y(0) e^{F\cdot Y}= L_{YX} (F) \widehat{R}_X(0) e^{F\cdot X}
\label{eq:gdb}
\ee

We remark that the stationary solution $\widehat{R}_X (F)$ does not
satisfy a detailed balance condition:

\be
L_{XY}(F) \widehat{R}_Y (F) \not= L_{YX}(F) \widehat{R}_X (F)
\label{eq:ndb}
\ee

\noindent An equality in eq.~(\ref{eq:ndb}) would imply that 
$\widehat{R}_Y (F)$  could be written as:

\be
\widehat{R}_X (F) \propto \widehat{R}_X(0)e^{F\cdot X}.
\ee

\noindent which is, evidently, not periodic.

In a local thermodynamic equilibrium the generalized stationary currents can 
be written in the following form:

\be
\widehat{J}_{x_\alpha} = \sum\limits_{\beta =1}^d \lambda_{\alpha \beta}
f_\beta.
\ee

The coefficients $\lambda_{\alpha \beta}$ are called Onsager coefficients. 
In general, even when the linear approximation cannot be applied, we can 
define:

\be
\lambda_{\alpha \beta} = \frac{\partial \widehat{J}_{x_\alpha}}{\partial f_\beta}.
\ee

We show in appendix~\ref{app:ons} that, provided the generalized detailed 
balance condition, eq.~(\ref{eq:gdb}), holds, these coefficients verify the 
generalized Gyarmati-Li~\cite{gyarli} reciprocal relations:

\be
\lambda_{\alpha \beta} = \lambda_{\beta \alpha} \quad \forall \, \alpha, \beta
\, \in \{ 1, \ldots , d \} 
\label{eq:ons}
\ee

\noindent and, hence, the Onsager reciprocity relations.
These properties are general and do not depend on the specific parameters 
of the model and are based on the generalized detailed balance condition
eq.~(\ref{eq:gdb}). This condition is a common assumption also for continuous 
models, as discussed further in the next section.

\section{Continuum limit for molecular motors}
\label{sec:cl}

\noindent

The usual choice for molecular motors is a 2--dimen-sional manifold in which 
one direction represents the position of the center of mass along the linear 
track (the microtubule or the actin filament). The other is the reaction 
coordinate for the $ATP$ hydrolysis, (see~\cite{kel00}), which is also related 
to the conformational changes of the motor protein.
These conformational changes are commonly thought to occur after binding 
$ATP$ and release of reaction products $ADP$ and $P_i$~\cite{hou00}.

The transition rates will be therefore written as $W_{xy}^{nm}$ where the 
subscript $xy$ denotes a transition from spatial position $y$ to spatial 
position $x$ whereas 
the superscript $nm$ stands for a transition from a state with $m$ ATP 
molecules to a state with $n$ ones\footnote{Using the definitions of the
previous section the spatial and chemical directions correspond to
$\alpha =1$ and $\alpha =2$ respectively.}. Of course this variable may also 
represent a non-integer chemical reaction coordinate (accounting for multiple 
states models), but this is the simplest possible choice. 
The periodicity is not specified for the spatial direction, while it is $1$ 
for the chemical direction, i.e. we are assuming that the state of the motor in
presence of $n$ ATP molecules is equivalent to the one in presence of $n+1$ ATP 
molecules. We remark that this assumption does not mean that a thermodynamic 
system with $n$ ATP molecules is equivalent to the same thermodynamic system 
with $n+1$ ATP molecules, but only that the chemical state of the motor protein 
after the reaction cycle is completed (and 1 ATP molecule is consumed or 
produced) is equivalent to the state it was before entering the cycle. 
We allow only transitions from a position $x$ to $x+a$ and $x-a$. All other 
transition rates will be identically $0$. 
Chemical transitions, i.e. those involving the superscripts $nm$ (with 
$n \ne m$) will be specified later. At the moment we only use transition 
rates in the form\footnote{The dependence on $n$ is
no more necessary, since $W_{xy}^{nm}$ is periodic in $n$ and $m$ with period
1.} $w_{xy}=\sum_m W_{xy}^{nm}$.
According to the definition eq.~(\ref{eq:defr}) and due to our choice of 
periodicity 1 in the chemical reaction coordinate, the 
master equation for the rate of change of $R_x$, or eq.~(\ref{eq:te1}) is: 

\bea
\dot{R_x} = w_{x,x+a} R_{x+a} & - & w_{x+a,x}
R_x + \nonumber \\ 
& + & w_{x,x-a} R_{x-a} - w_{x-a,x} R_x.
\label{eq:ev_mot}
\eea

%

We define a discrete current $j_x$:

\be
j_x = \left( w_{x+a,x} R_x - w_{x,x+a} R_{x+a} \right),
\label{eq:defdc}
\ee

Applying this definition to  eq.~(\ref{eq:ev_mot}):

\be
\dot{R_x} = - \left(j_x - j_{x-a}\right) \equiv - a (\nabla_d j)_x
\label{eq:defjd}
\ee

\noindent 
where $(\nabla_d j)_x$ is by definition the discrete gradient of $j_x$.

In the continuum (FP) description:

\bea
\dot{P}(x,t) & = & -\nabla j(x,t) \label{eq:FP1}, \\
j(x,t) & = & \frac{D}{T} \left[- T \nabla P(x,t) - P(x,t) \nabla V(x) + P(x,t) f
\right] \nonumber \\ 
& & \label{eq:FP2}
\eea

\noindent where $T$ is the temperature, $D$ the diffusion constant, $V(x)$ a 
periodic potential, $f$ the force and $j(x,t)$ is the probability density 
current. The potential $V(x)$ has the same period as the transition rates 
$w_{xy}$.

%

For very small values of $a$ we define:

\be
R_x \equiv a P(x,t)
\label{eq:defcont}
\ee

If in eq.~(\ref{eq:defdc}) we expand $R_{x+a}$ to the first order in $a$, the
discrete current can be rewritten as:

\be
j_x = R_x (w_{x+a,x}-w_{x,x+a}) - a (\nabla R_x) w_{x,x+a}
\label{eq:J}
\ee

Using eqs.~(\ref{eq:defjd}), (\ref{eq:FP1}) and (\ref{eq:defcont}), for very
small $a$, one gets:

\be
j_x = j(x,t).
\label{eq:corresp}
\ee

By means of this correspondence, eq.~(\ref{eq:corresp}), and using
eqs.~(\ref{eq:FP2}) and (\ref{eq:J}), we obtain:

\bea
D & = &  a^2 w_{x,x+a} \label{eq:sol1}\\
-D \nabla V(x) + D f & = & a \left( w_{x+a,x}-w_{x,x+a} \right)
\eea

Solving for $w_{x+a,x}$:

\be
w_{x+a,x}= w_{x,x+a} [ 1 - a \beta (\nabla V(x) - f) ],
\label{eq:wxax}
\ee

\noindent where eq.~(\ref{eq:sol1}) has been used. If we suppose that the 
following relation holds:

\be
w_{x+a,x} = w_{x,x+a} e^{-\beta (V(x+a) - V(x) - a f)},
\label{eq:dbf}
\ee

\noindent then, eq.~(\ref{eq:wxax}) is satisfied in the continuum limit ($a \to
0$).

This relation corresponds to the standard detailed balance condition 
when $f=0$ and the stationary periodic solution is:

\be
\widehat{R}_x(0) \propto e^{-\beta V(x)},
\ee

\noindent whereas eq.~(\ref{eq:dbf}) corresponds to a generalized detailed balance 
condition, eq.~(\ref{eq:gdb}), when $f \not= 0$.

\subsection{The chemical reaction}

By analogy with the continuum model~\cite{jap97}, we suppose that 
each transition rate $W_{xy}^{nm}$ is essentially due to 3 subprocesses, 
leading to ATP consumption ($\alpha $ transitions in \cite{par99}), 
ATP production ($\gamma$) and no change in ATP concentration or thermal 
transitions ($\beta$).

Since a chemical reaction is present in $\alpha $ and $\gamma$ processes, the 
chemical potential difference, $\Delta \mu = \mu_{ATP}-\mu_{ADP}-\mu_{P_i}$, 
plays the role of a generalized force, conjugated to the number of ATP 
molecules. In this paper only transitions from $m=n+\Delta n$ to $n$ ATP
molecules with $\Delta n = \pm 1, 0$ will be considered.

The generalized detailed balance condition eq. (\ref{eq:gdb}) is therefore:

\be
\frac{W_{x+a,x}^{n, n + \Delta n}}{W_{x,x+a}^{n + \Delta n,n}} =
 e^{-\beta (V(x+a)-V(x)-f a -\Delta n \Delta \mu)}
\label{eq:dbalfa}
\ee

where $e^{-\beta V(x)} \propto \widehat{R}_x (0)$ is the stationary
equilibrium solution when all generalized forces are zero.
Transitions with $\Delta n > 0 (< 0)$ correspond to ATP consumption
(production).
Eq.~(\ref{eq:dbalfa}) states that when $\Delta \mu > 0$ ($\Delta \mu < 0$)
transitions leading to ATP consumption (production) are more favorable and 
lead to a spatial advancement of the motor. This is the core of the energy 
transduction process: chemical energy is used to perform mechanical 
work against a load $-f$ or chemical energy is produced performing a mechanical
work on the protein. Notice that, according to our choice of notation, 
$f_1= \beta f$ and $f_2=-\beta \Delta \mu$.

These generalized detailed balance conditions are the same introduced in
\cite{par99}, in the limit $a \to 0$. We remark that this scheme corresponds to 
a periodicity $1$ in the chemical reaction coordinate. A different
periodicity in the chemical coordinate would allow to introduce different 
potential shapes $V (x, \xi )$ for each value of the reaction coordinate 
$\xi$, as for instance in~\cite{lip00}. Using periodicity 2 in the chemical
coordinate, the continuous two--state model \cite{jap97,par99} is fully recovered 
in the $a\to 0$ limit with the same detailed balance conditions as in
\cite{jap97,par99} \footnote{Indeed, using periodicity $2$ for the chemical
coordinate, if we allow only transitions between
neighboring sites both in the spatial and chemical coordinate,
eq.~(\ref{eq:te1}) becomes:
\bea
\dot{R_x^1} & = & W^{11}_{x,x+a} R^1_{x+a} - W^{11}_{x+a,x} R^1_x + W^{11}_{x,x-a} 
R^1_{x-a} + \nonumber \\ 
       & - & W^{11}_{x-a,x} R^1_x + W^{12}_{xx} R_x^2 - W^{21}_{xx} R_x^1 
\label{eq:note1} \\
\dot{R_x^2} & = & W^{22}_{x,x+a} R^2_{x+a} - W^{22}_{x+a,x} R^2_x + W^{22}_{x,x-a} 
R^2_{x-a} + \nonumber \\ & - & W^{22}_{x-a,x} R^2_x + W^{21}_{xx} R_x^1 
- W^{12}_{xx} R_x^2
\label{eq:note2}
\eea

where $1$ and $2$ represent the two conformational state indexes and not the
number of ATP molecules. In the continuum limit, this model is perfectly 
equivalent to the ones proposed in~\cite{jap97,par99}, with the following 
substitutions: $w_{x,x\pm a} = W^{ii}_{x,x\pm a}$, $V(x)=V_i(x)$, $i=1,2$ 
in eq.~(\ref{eq:dbf}) and $W^{12}_{xx}=\omega_2(x)$, $W^{21}_{xx}=
\omega_1(x)$ in eqs.~(\ref{eq:note1}--\ref{eq:note2}) in the notation 
of refs.~\cite{jap97,par99}. The effective mobility $\xi^{-1}$ of 
refs.~\cite{jap97,par99} is $\beta a^2 W_{x,x+a}^{ii}$.}.
The general scheme introduced here allows to treat the chemical and mechanical
coordinates on an equal footing and write the detailed balance condition in a
more standard and transparent way.

\subsection{Generalized currents}

According to the definitions given in section \ref{sec:gf}, the stationary 
current $\widehat{J}_t$ ($t$ stays for translational motion), associated to 
the protein center of mass corresponds to the velocity of the motor protein. 
Its full expression is given by:

\be
\widehat{J}_t = a \sum\nolimits'_x \left[ w_{x+a,x} -w_{x-a,x} \right] \widehat{R}_x .
\label{eq:vel},
\ee 

\noindent whereas the stationary current $\widehat{J}_c$ associated to the chemical
coordinate is:

\be
\widehat{J}_c = \sum\nolimits'_x \left( W_{x+a,x}^{n,n+1} + W_{x-a,x}^{n,n+1}- 
W_{x+a,x}^{n,n-1} - W_{x-a,x}^{n,n-1}  \right) \widehat{R}_x
\label{eq:atpprod},
\ee

\noindent which does not depend on $n$ since $W_{xy}^{nm}=W_{xy}^{n+1,m+1}$.

This current corresponds to the number of ATP mole-cules consumed per unit time,
so we will refer to it as the ``rate of ATP consumption''. 
Notice that this current has the opposite sign of the one in eq.(19).
This is consistent with the above choice $f_2=-\beta \Delta \mu$. We
remark that  $\beta $ transitions do not contribute to ATP production since
they do not  involve any chemical reaction. 
These considerations will be used in the next section to develop and fully
characterize a very simple discrete model whose continuum limit is still
described by a Fokker--Planck equation. 

\section{Definition of the model}
\label{sec:def}

We showed that eq.~(\ref{eq:dbf}), i.e. a generalized detailed balance condition, 
is sufficient for a discrete model to be compatible with a Fokker--Planck 
equation in the continuum limit. 
Actually eq.~(\ref{eq:dbf}) is compatible with a very general class of 
models, for which any apportionment of force $f$ and chemical potential 
difference $\Delta \mu$ over forward (in space) and backward transitions, is 
perfectly reasonable.
This is true as long as each transition allows the protein to take a substep
which can be made infinitesimally small. If this is not possible for
some of these substeps, the simple Fokker--Planck equation may not be able
to describe the system in the continuum limit.
Nevertheless the Onsager relations still hold, as long as eq.~(\ref{eq:gdb}) 
holds and provided that the sum of all substeps is equal to one period. 
The continuum limit, left alone, is therefore not sufficient to fix the
parameters required to define discrete models.

To be as general as possible, we use forward  ($u_j$) and backward ($w_j$) 
transition rates, with a priori not specified apportionments:

\bea
u_j(f,\Delta \mu ) & = & \omega_j (f, \Delta \mu) e^{\beta (-v_j+a_j^+ f + m_j^+ 
\Delta \mu)} \\
w_j (f,\Delta \mu ) & = & \omega_j (f, \Delta \mu) e^{\beta(- a_j^- f - m_j^-
\Delta \mu)},
\eea
\noindent
where $v_j$ is the potential
difference between states $j+1$ and $j$ while $\omega_j(0,0)$ corresponds to a
spontaneous transition  probability from state $j$ to state $j+1$ in the
absence of any potential difference and generalized force.
Since the potential is periodic, the $v_j$'s are subject to the condition 
$\sum\nolimits_j v_j = 0$.
In the absence of any potential difference and generalized force, the particle
can diffuse in both directions, so $u_j(0,0)=w_j(0,0)=\omega_j(0,0)$.
This choice corresponds to a FP equation in two dimensions where both the
spatial and chemical coordinates, as well as their associated generalized 
forces, are treated on an equal footing\footnote{This equation may be written 
in the following form:

\bea
& & \frac{\partial P(x,\xi,t)}{\partial t} = - \frac{\partial 
J_1(x,\xi,t)}{\partial x} -
\frac{\partial J_2(x,\xi,t)}{\partial \xi} \\
& & J_1 = D_1 \left[- T \frac{\partial P}{\partial x} - P 
\frac{\partial V(x,\xi)}{\partial x} + P f \right] \\
& & J_2 = D_2 \left[- T \frac{\partial P}{\partial \xi} - P 
\frac{\partial V(x,\xi)}{\partial \xi} + P \Delta \mu \right]
\eea

where $x$ and $\xi$ are, respectively, the spatial and chemical coordinates.
}.

According to eq.~(\ref{eq:gdb}), all sums $a_j^++a_j^-$ ($m_j^++m_j^-$) should 
be interpreted as the effective size of the spatial (chemical) substep taken 
by the motor protein. Some recent experiments~\cite{nish01} seem to suggest 
the existence of these small substeps.
A substep can be of different types: only positional ($m_j^++m_j^-=0$), only
chemical ($a_j^++a_j^-=0$) or mixed. The $f$ and $\Delta \mu $ dependence 
in the transition rates, $u_j$'s and $w_j$'s, accounts also for more 
complicated schemes which cannot be ruled out, in principle.

The substep size is unknown and it is related to the conformational changes 
involved after binding ATP, but on a pure theoretical basis, space may be 
discretized so as to obtain the largest unit of substep such that all 
``natural'' substeps are multiple of this elementary unit. 
After all $N$ substeps, a full spatial and chemical period has been covered, 
so that:

\be
\frac{a_j^++a_j^-}{p}= m_j^++m_j^-=\frac{1}{N},
\label{eq:steps}
\ee

\noindent where $p$ is the typical spatial step performed by the motor. We 
assume that the $\omega_j$ for this elementary unit
of substep do not depend on $f$ and $\Delta \mu$. This hypothesis is commonly
assumed in discrete models for substeps of any size~\cite{fish99,lip00,ast96}, 
whereas in our model it is assumed only for these elementary substeps. For a
natural substep a more complicated force dependence of transition rates is
possible, at least in principle.

At the same time, since in absence of any internal potential or external
generalized force the probability to proceed in one direction or the other is
simply given by the probability to diffuse by $1/N$ of a full period in both
coordinates, we assume that $\omega_j$ does not depend on $j$ and omit the 
subscript in the transition rates $\omega_j$.

It is possible to show that such a model, in general, is not compatible with 
very simple requests, i.e. that the velocity of the motor saturates for high 
ATP concentrations, and that the velocity obeys a Michaelis law, as follows 
from the data of~\cite{lat01,viss99}.
Indeed, using eq.~(\ref{eq:steps}) it is possible to show that the velocity 
is given in general by~(see~\cite{der83}):

\be
J_t=\frac{p\omega\left(e^{\beta\Delta\mu} -1 \right)}{{\mathcal{S}}}
\ee

\noindent where

\bea
{\mathcal{S}} & = & \left. \sum\limits_{n=1}^{N}
e^{\beta[v_n+(1-m_n^+)\Delta \mu]} \right[ 1+ \nonumber \\
& + & \left. \sum\limits_{i=1}^{N-1}
e^{-\beta\left(\frac{i-1}{N}+m_{n+i}^++m_n^-\right)\Delta \mu} 
\prod\limits_{j=1}^i e^{\beta v_{n+j}} \right]
\eea

\noindent in the case where $f=0$. If all $m_n^\pm \ge 0$, in the large $\Delta 
\mu$ limit the velocity is given by:

\be
J_t \approx \frac{p\omega}{\sum\nolimits_n e^{\beta(v_n-m_n^+\Delta \mu)}}
\ee

\noindent which is exponentially large in $\Delta \mu$ unless all $m_n^+=0$. 
But this condition, together with the condition that $m_n^++m_n^-=\frac{1}{N}$ 
implies that the velocity can be written in the form ($q=e^{\beta \Delta \mu}$):

\be
J_t =\frac{A (q-1)}{K_N q + K_{N-1}q^{\frac{N-1}{N}}+\ldots + K_1
q^{\frac{1}{N}}} \quad , \quad q = e^{\beta \Delta \mu}
\label{eq:notmich},
\ee

where the constants $K$'s do not depend on $q$ and are model dependent.
Eq.~(\ref{eq:notmich}) is not a Michaelis law, i.e. of the form 
\mbox{$J_t = \frac{Aq}{K_M +q}-B$}, even in the large $q$ limit. This is true for 
any value of $N$ except $N=1$. Therefore one such discrete model is compatible 
with a Michaelis law only if we concentrate all the chemical reaction in a single 
substep. Thus let us assume now that the chemical reaction occurs in a
single step and  without loss of generality we suppose that this ATP driven
step is the first one  in the cycle. 
We notice that this in turn would imply that the smallest substep is the one 
which occurs during ATP hydrolysis, which is reasonable\footnote{A net
advancement of the protein center of mass is commonly thought to occur upon
release of the reaction products, whereas the hydrolysis reaction does not 
seem to imply any net macroscopic rearrangement on its own~\cite{rice99}.}.

\begin{figure}
\begin{center}
\epsfig{figure=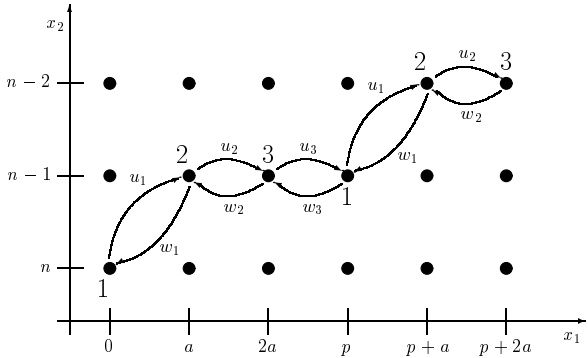,scale=0.5,clip=}
\caption{\label{fig:disc} Graphical representation of a discrete model, with
$N=3$ states along the spatial direction $x_1$. The chemical direction $x_2$
represents the number of ATP molecules. Conformational states of the motor
protein are represented by the numbers $1,2,3$. The system is periodic both 
in the spatial direction $x_1$ (periodicity $p$; $a=p/N=p/3$) and in the 
chemical direction (periodicity $1$). Only transition rates $u_1$ and $w_1$ 
involve a chemical reaction, all the others are purely positional. }
\end{center}
\end{figure}

A graphical representation of this class of discrete models with only one
chemical transition is given in fig.~\ref{fig:disc}, for the case where $N=3$.

Let $\Delta \mu$  be apportioned over the forward and backward transitions, so 
that $m_1^+=\epsilon, m_1^-=1-\epsilon$. With some calculations it is possible to 
show that in this case we obtain for velocity an expression of the type:

\be
J_t=\frac{A(q-1)}{Kq+K_1q^\epsilon+K_2q^{1-\epsilon}},
\ee

\noindent which is still incompatible with a Michaelis law unless $\epsilon=0$ 
or $\epsilon=1$. Choosing one or the other leads essentially to the same 
Michaelis law, except some multiplicative parameters which in turn depend on 
the products $\omega e^{\beta v_j}$, and ought to be determined by experiments.

In the same way, the force apportionment can be investigated. The problem is
that at variance of chemical potential, experimental data are obtained only for
small forces. Therefore it is not possible to use the same criteria since,
to our knowledge, for high values of force the velocity may grow up even to
$\pm \infty$, meaning that the motor detaches from the fiber. Nevertheless,
by the same line of reasoning, if we assume that the velocity should reach a
constant value for high values of force in the positive direction, we find the
condition:

\be
a_j^+=0 \quad \forall j=1,\ldots, N.
\ee

On the contrary, if we assume that a constant velocity is reached only when the
external force opposes the natural direction of movement (which is what one
expects on the basis of the available data~\cite{viss99}), then we find the
condition that:

\be
a_j^-=0 \quad \forall j=1,\ldots, N
\ee

All intermediate cases, the symmetric one included, lead, therefore, to 
velocities growing up to $\pm \infty $ as the force tends to $\pm \infty$. 
Remarkably, this is what happens when using a FP equation with very high 
forces, since the probability distribution is not sensitive to the potential 
shape and becomes flat, i.e. velocity is a linear function of force.
Of course in a FP description every apportionment is totally equivalent to any 
other, as long as we consider the small $a$ limit. This is not true in the 
limit for very high forces, for which our derivation of the continuum limit 
is no longer a good approximation. 
Therefore it is not surprising that in discrete models 
a particular choice of apportionment has dramatic consequences on the 
asymptotic behaviors of the quantities of interest, in the same way a 
different apportionment of the chemical potential leads to asymptotic 
behaviors which are not compatible with the expected Michaelis law unless, as
shown above, the chemical reaction is concentrated in a single step. 
In the following section we will study in detail some examples of models with
spatial periodicity 2 and 3 highlighting their main predictions.
 
\section{Model predictions and example studies}
\label{sec:mpaes}

In all the following, lengths are measured in units of filament
periods $p$, while potential difference, $v_j$, and chemical potentials,
$\Delta \mu$,  are measured in units of $kT$, forces, $f$, in units of $kT/p$. 

According to our previous discussion, since experimental data suggest that the
velocity should reach a constant value for high negative forces, we assume the
force apportionment to be asymmetric and concentrated in the forward
transitions. In the same way we assume the first step to be chemically driven,
so as to obtain a Michaelis law. We will show that these two hypotheses, left
alone, are sufficient to obtain a force dependent Michaelis constant for 
velocity and rate of ATP consumption, and also some interesting predictions 
about effective step--size and randomness factor.

In these models:

\bea
u_n & = & \omega e^{-v_n+\frac{f}{N}+\Delta \mu \delta_{n1}} \label{eq:tr1} \\
w_n & = & \omega \label{eq:tr2},
\eea

\noindent with the condition $\sum_n v_n = 0$. The expression for velocity, using 
eqs.~(\ref{eq:tr1}) and (\ref{eq:tr2}), is:

\bea
J_t & =& \frac{p\omega\left(e^{\beta (\Delta \mu + f)} -1 
\right)}{{\mathcal{S}}} \label{eq:velex}\\
{\mathcal{S}} & = & \left. \sum\limits_{n=1}^{N} e^{v_n+(1-\delta_{n1})\Delta 
\mu+(1-\frac{1}{N})f}\right[ 1+ \nonumber \\
& + & \left. \sum\limits_{i=1}^{N-1} e^{-\left( 
\frac{i}{N}\right)f} \prod\limits_{j=1}^i
e^{v_{n+j}-\delta_{n+j,1}\Delta \mu}\right].
\eea

Interestingly, the correct law for velocity is of the type:

\be
J_t=\frac{Aq}{K_M+q}-B,
\label{eq:michlatt}
\ee

\noindent where $K_M$ is the Michaelis constant. The same result has also been obtained
in~\cite{lat01} in the context of continuous models. 
Eq.~(\ref{eq:velex}) may be used to calculate the `stall' force, i.e. the force
for which the velocity is zero. This is simply given by:

\be
f_{stall}=-\Delta \mu.
\label{eq:stall}
\ee

This picture is, of course, very simplified. The experimental data reposted 
in~\cite{viss99} show that the stall load ($-f$) increases with increasing ATP 
concentrations, but probably the dependence on $\Delta \mu $ is not as simple as
in eq.~(\ref{eq:stall}).
However data in this parameter region are subject to large experimental errors, 
due to rapid detachment of beads from the microtubule under stall conditions, 
so a comparison with experiment, at present, can be only of qualitative nature. 

The rate of ATP consumption follows from eq.~(\ref{eq:atpprod}) and can be
written as:

\be
J_c=\frac{u_1 R_1 - w_1 R_2}{\sum\nolimits_n R_n}.
\label{eq:jc}
\ee

Using eq.~(\ref{eq:velex}) and eq.~(\ref{eq:jc}), it is possible to show that
the rate of ATP consumption and the velocity, when measured in ATP molecules 
hydrolyzed per second and periods per second respectively, are exactly the 
same quantity, so that the effective step size (the number of periods taken 
per hydrolyzed ATP molecule) is $1$, which is consistent with experiments on
kinesin~\cite{schn97}.
This is no longer true if we use a more complicated scheme with pure thermal
processes in addition to the normal ATP consuming ones, leading to transitions 
between different states: in this case the effective step size can be
smaller  than one.

Using eq.~(\ref{eq:velex}) it is possible to obtain an explicit expression for 
the force dependent Michaelis constant; for simplicity we concentrate on models
with spatial periodicity $2$ (2--models) and $3$ (3--models). For 2--models
$v_1=v,v_2=-v$, whereas for 3--models we assume $v_1=v,v_2=-v/2, v_3=-v/2$. 
This means that the chemical potential difference is used in the first 
transition to overcome the internal potential barrier $v$, while the other 
transitions are favored by the internal potential shape. The Michaelis constant
of eq.~(\ref{eq:michlatt}) is given by:

\bea
K_M(2) &=&  e^{2v}+2 e^{-\frac{f}{2}+v} \\
K_M(3) &=& \frac{2e^{\frac{3}{2}v}+3 e^{-\frac{f}{3}+v}+e^{\frac{f}{3}+2v}}{1+2
e^{\frac{f}{3}+\frac{v}{2}}},
\eea

\noindent for 2-- and 3--models respectively.
In both examples, the Michaelis constant grows exponentially with $-f$ (see
discussion of section~\ref{sec:def}), in accordance with the recent experimental 
observation of an increase in the Michaelis constant with applied 
load~\cite{viss99}. Interestingly both models predict a constant value for 
$K_M$ at high positive values of the force. We remark that experimental 
data on the Michaelis constant for positive forces, to our knowledge, are 
still unavailable.

Recently developed experimental techniques allowed to measure the
randomness parameter~\cite{svo94}, defined as the long time limit of the ratio 
between the variance of the protein position on the filament,
$<x^2(t)>-<x(t)>^2$, and the product of its average position, $<x(t)>$ and 
periodicity $p$:

\be
r=\lim_{t \to \infty} \frac{<x^2(t)>-<x(t)>^2}{<x(t)>p}
\ee

In facts, the time resolution of experiments corresponds to the same order of
magnitude of one hydrolysis event (typically milliseconds), while, at present, 
conformational changes leading to ATP hydrolysis cannot be directly measured. 
Nonetheless they affect the statistical distribution of the protein 
movement and the randomness parameter \footnote{See~\cite{wang98} for a 
very interesting discussion on the relevance of the randomness parameter and 
the effective diffusion constant for mechanochemical transducers.}. 
The macroscopic diffusion coefficient $D$
is defined so that: $<x^2(t)>-<x(t)>^2 = 2 D t $ , while $<x(t)> = V t$, where 
$V$ is the velocity. Therefore:

\be
r=\frac{2D}{pV}.
\label{eq:rand}
\ee

The expression for $D$ is rather complicated, but still calculable for jump 
processes~\cite{der83}.
The randomness parameter presents some interesting
properties~\cite{viss99,svo94,par00}: first, $r=0$ for a perfectly clocklike
motor, whereas $r=1$ for a `Poisson' motor with one biochemical transition and
exponentially distributed time intervals between events. In general $r^{-1}$
provides a continuous measure of the number of rate-limiting transitions in the
overall mechanochemical cycle (see~\cite{par00} for a simple proof in the case
of a jump process with equal forward and backward transition rates).
In our models the expressions for the randomness parameter can be easily
calculated, but their expressions are rather cumbersome.
Some important features can be directly inferred and compared with
experiments.

At small loads $-f$ and small $\Delta \mu$ the velocity decreases linearly with 
both $-f$ and $\Delta \mu$, so that the randomness factor from eq.~(\ref{eq:rand}) 
should approach $\infty$, assuming that the macroscopic diffusion coefficient 
approaches a constant value, in general different from zero. This should be 
also verified under stall conditions, when the randomness factor again 
approaches $\infty $. 

\begin{figure}
\begin{center} 
\epsfig{figure=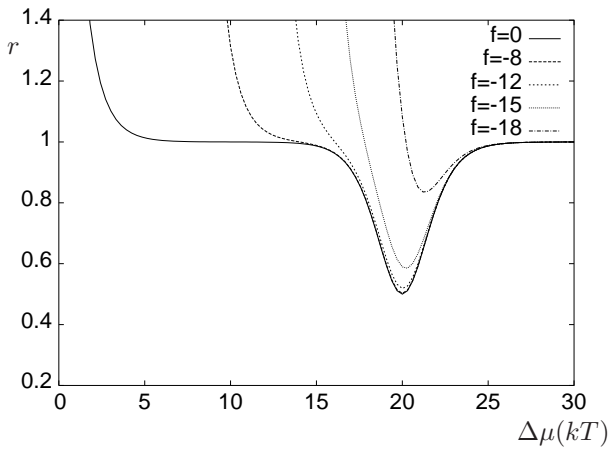,scale=0.3,clip=,angle=270}
\put(-225,-18){ $ r $ } 
\put(-32,-164){ $\Delta \mu (kT)$}
\end{center}
\caption{\label{fig:r2} Randomness parameter in a 2--model versus $\Delta \mu$, 
with $v=10\ kT$. Randomness approaches the value $1/2$ when $\Delta \mu \approx 2v$,
indicating that both transitions are rate limiting. At high $\Delta \mu $ there
is only one rate limiting transition, so that randomness approaches $1$.
The force $f$ is  measured in units of $kT/p$.}
\end{figure}

\begin{figure} 
\begin{center} 
\epsfig{figure=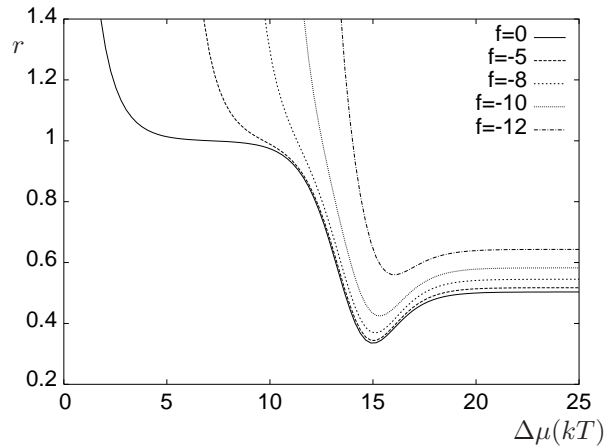,scale=0.3,clip=,angle=270}
\put(-225,-18){ $ r $ } 
\put(-32,-164){$\Delta \mu (kT)$}
\end{center}
\caption{\label{fig:r3} Randomness parameter in a 3--model versus $\Delta \mu$,
with $v=10\ kT$. Randomness approaches the value $1/3$ when $\Delta \mu \approx
3v/2$, indicating that all transition rates are rate limiting. At high $\Delta
\mu $, there are essentially two rate limiting transition rates, so that
randomness approaches $1/2$, at least at low loads.
The force $f$ is  measured in units of $kT/p$.}
\end{figure}

At low values of $\Delta \mu $, but still sufficient to force the protein out of
the stall condition, the randomness factor exhibits different 
behaviors, depending on force, as shown in figures~\ref{fig:r2} and~\ref{fig:r3}.
When $\Delta \mu < v$, the rate limiting step is essentially the first
one for both models, so the randomness factor approaches 1 when 
the stall condition is overcome. At intermediate values, $v < \Delta \mu 
< 2v$ for 2--models and $v < \Delta \mu < \frac{3}{2}v$ 
for 3--models, the rate limiting transition is still the first one, but its
rate limiting power is decreasing with respect to the others, due to the
increase in $\Delta \mu$, so that the randomness parameter is decreasing, until 
at the border of this parameter region, $\Delta \mu\simeq v$ 
for 2-- and $\Delta \mu\simeq \frac{3}{2}v$ for 3--models, all forward 
transition rates have essentially the same value: all steps are equally 
rate limiting, so that there are $2$ rate limiting steps for 2--models and 
$3$ for 3--models. This is evident also in the figures; in this part of the 
graphs, the randomness parameter approaches $1/2$ and $1/3$ for 2-- and 
3--models respectively.

At very high ATP concentrations, the first step has a high occurrence
probability, whereas the other steps are rate limiting. This implies that the
randomness parameter should approach a constant value $1$ for 2--models and
$1/2$ for 3--models. All these behaviors may be observed in the figures, at
least for small values of force.  Under very high loads, the velocity of the
motor protein is very low in a wide range of $\Delta \mu$ values. This implies
that the randomness factor is very high, until $\Delta \mu$ is high enough to
force the system out of the stall condition.
A direct comparison with experiments would be
desirable. The central part of our figures reproduces data obtained for
randomness in~\cite{viss99}. Our  predictions in this parameter region also
agree with theoretical derivations  for continuous two-state ratchet models on
kinesin~\cite{par00}, but in the case of continuous models it is very
difficult to calculate the randomness parameter at very high ATP
concentrations, due to numerical problems. This is no longer true for discrete
models, for which all complications are algebraic, rather than numerical. 
The experimental measure of randomness under very small loads or for small 
positive values of force should also be useful to infer, on a quantitative basis, 
the spatial periodicity necessary to fully characterize the system. Data 
in~\cite{viss99} seem to suggest a 3--model, since the randomness factor 
approaches a minimum value close to $1/3$, but measurements 
at smaller loads would be useful to confirm or reject this hypothesis.

\section{Conclusions}

In this paper, we have introduced a master equation approach to
describe interesting properties of molecular motors. Our approach is similar in
spirit to a Kramers--Moyal expansion of the FP equation~\cite{risken}, but it is
specifically studied for the problem of motor proteins, where both the
mechanical and chemical coordinates are important.
In this framework we have studied the general conditions needed to obtain a 
discrete chemical kinetics model from a continuous one, when all generalized 
coordinates are treated on an equal footing. In this limit, a FP equation is
recovered, when the requirement of a generalized detailed balance condition, 
eq.~(\ref{eq:gdb}), is fulfilled in the out of equilibrium regime. This 
condition has been shown to imply the validity of the Onsager reciprocity 
relations for the periodic stationary solutions in the 
close to equilibrium regime. This property is not surprising, since both 
Onsager relations and detailed balance are supposed to descend from microscopic 
reversibility. Motor proteins operate far from equilibrium, and far from the
linear regime, where reciprocity relations are expected to hold. Nevertheless
reciprocity relations constitute a minimal requirement in a model for
mechanochemical transduction, as firstly stated by Hill~\cite{hill74} and, more
recently, in the context of continuous models~\cite{par99}. Moreover, to our
knowledge, a proof for the case of discrete models, has never been given.

The continuum limit has been shown to be insufficient to fix all parameters in 
a discrete chemical kinetics model for any cyclic out of equilibrium 
thermodynamic system. 
However, in the case of motor proteins, semi--phenomenological considerations 
led us to the formulation of discrete models which are compatible with the
stochastic continuous ones in the continuum limit, satisfy the Onsager
reciprocity relations and allow an easy comparison with experimental data.

We believe this work to be useful to study the general conditions which ought to
be verified by a discrete kinetics model and, most importantly, to study the
force dependence of transition rates in mechano--chemical processes. This is of
relevance not only for research on motor proteins, but also on other important
biomolecules.

\acknowledgements
GL acknowledges P. V\'an for a critical reading of the manuscript.

\begin{appendix}

\section{Onsager coefficients and relations}
\label{app:ons}

In this appendix we will prove that, provided a generalized detailed balance
condition in the form~(\ref{eq:gdb}) holds, the Onsager relations
eq.~(\ref{eq:ons}) hold for any discrete model. We also provide a derivation of
the Onsager coefficients and give their values for the case of 2-- and
3--models\footnote{In these simplified models $J_c=J_t$, but this does
not imply that the Onsager reciprocity relations are trivially satisfied. If
they are,  $J_c=J_t$ simply implies that all coefficients should be
equal.} studied in  section~\ref{sec:mpaes}.
In the following we use the notation $\partial_\alpha$ to denote a partial 
derivative with respect to a generalized force $f_\alpha$.

Differentiating both sides of eq.~(\ref{eq:gdb}), we obtain the condition:

\bea
& & \left[ \left( \partial_\alpha L_{XY}(F) \right) \widehat{R}_Y (0) - \left( 
\partial_\alpha L_{YX}(F) \right) \widehat{R}_X (0) \right]_{F=0}  = \nonumber \\
& = & L_{XY}(0) \widehat{R}_Y (0) (x_\alpha - y_\alpha).
\label{eq:diffgdb}
\eea

Differentiating the stationarity condition, eq.~(\ref{eq:st}), and applying
eq.~(\ref{eq:diffgdb}) we obtain, after some manipulation:

\bea
& & \left[ \sum\nolimits_Y  L_{XY}(F) \partial_\alpha
R_Y(F)\right]_{F=0} = \nonumber \\ 
& & \ \ \ \ \ \ \ \ \ \ \ \ \ \ \ \ \ \ \ = \sum\nolimits_Y L_{XY}(0) \widehat{R}_Y (0) 
(y_\alpha - x_\alpha).
\label{eq:diffst}
\eea

If $A_{XY}=A_{X+LN,Y+LN} \, \forall L \, \in \, {\cal{Z}}^d$:

\bea
& & \sum\nolimits_{XY} \chi_{_X} A_{XY} = \sum\nolimits_{XY} \chi_{_Y} A_{XY} =
\nonumber \\
& = &\sum\nolimits_{XY} \frac{\chi_{_X}+\chi_{_Y}}{2} A_{XY},
\label{eq:lastprop}
\eea 

\noindent as a consequence of eq.~(\ref{eq:prop}).

Replacing $q_X$ with $x_\alpha$ in eq.~(\ref{eq:symj}) and differentiating with
respect to $f_\beta $, we obtain:

\bea
& & \lambda_{\alpha \beta} =  \sum\nolimits_{XY} \frac{\chi_{_X} + \chi_{_Y}}{4} \left( x_\alpha - y_\alpha \right)
\left[ \left( \partial_\beta L_{XY} \right) \widehat{R}_Y + \right. \nonumber \\
& & - \left.\left( \partial_\beta L_{YX} 
\right) \widehat{R}_X + L_{XY} ( \partial_\beta \widehat{R}_Y ) -  L_{YX}
( \partial_\beta \widehat{R}_X ) \right]_{F=0}. \nonumber \\
& &  
\eea

Using condition~(\ref{eq:diffgdb}), we are left with:

\bea
& & \lambda_{\alpha \beta} = \nonumber \\
& & = \sum\nolimits_{XY} \frac{\chi_{_X} + \chi_{_Y}}{4} 
\left( x_\alpha - y_\alpha \right) \left( x_\beta - y_\beta \right)
L_{XY}(0) \widehat{R}_Y(0) + \nonumber \\
& & + \sum\nolimits_{XY} \frac{\chi_{_X} + \chi_{_Y}}{4} \left( x_\alpha - y_\alpha
\right) \left[  L_{XY} ( \partial_\beta \widehat{R}_Y ) + \right. \nonumber \\ 
& & - \left. L_{YX} ( \partial_\beta \widehat{R}_X ) \right]_{F=0}.
\eea

Applying the property~(\ref{eq:lastprop}) to the quantities 
$(x_\alpha - y_\alpha ) L_{XY}(0) ( \partial_\beta \widehat{R}_Y )$ and 
$(x_\alpha - y_\alpha ) L_{YX}(0) ( \partial_\beta \widehat{R}_X )$ the
Onsager coefficients are:

\bea
& & \lambda_{\alpha \beta} = \nonumber \\
& & = \sum\nolimits_{XY} \frac{\chi_{_X} + \chi_{_Y}}{4} 
\left( x_\alpha - y_\alpha \right) \left( x_\beta - y_\beta \right)
L_{XY}(0) \widehat{R}_Y(0) + \nonumber \\
& & + \frac{1}{2} \sum\nolimits_{XY} \left( x_\alpha - y_\alpha \right) \left[ 
\chi_{_Y} L_{XY} (\partial_\beta \widehat{R}_Y) + \right. \nonumber \\
& & - \left. \chi_{_X} L_{YX} (\partial_\beta
\widehat{R}_X) \right]_{F=0} . 
\eea

A subsequent application of eq.~(\ref{eq:diffst}) and eq.~(\ref{eq:gdb}) leads 
to:

\bea
& & \lambda_{\alpha \beta} = \nonumber \\
& & = \sum\nolimits_{XY} \frac{\chi_{_X} + \chi_{_Y}}{4} 
\left( x_\alpha - y_\alpha \right) \left( x_\beta - y_\beta \right)
L_{XY}(0) \widehat{R}_Y(0) + \nonumber \\
& & + \frac{1}{2} \sum\nolimits_{XY} \left[ \chi_{_Y} \frac{L_{YX}}{\widehat{R}_Y}
( \partial_\alpha \widehat{R}_X) ( \partial_\beta \widehat{R}_Y) + \right.\nonumber \\ 
& & + \left. \chi_{_X}  
\frac{L_{XY}}{\widehat{R}_X} ( \partial_\alpha \widehat{R}_Y) ( \partial_\beta 
\widehat{R}_X) \right]_{F=0}.
\eea

From eq.~(\ref{eq:gdb}) and another application of eq.~(\ref{eq:lastprop}), we
finally obtain:

\bea
& & \lambda_{\alpha \beta} = \nonumber \\ 
& & =\sum\nolimits_{XY} \frac{\chi_{_X} + \chi_{_Y}}{4}
\left[ \left( x_\alpha - y_\alpha \right) \left( x_\beta - y_\beta \right)
L_{XY} (0) \widehat{R}_Y (0) + \right. \nonumber \\
& & + \left. \frac{L_{XY}(0)}{\widehat{R}_X(0)} A_{XY}(0) \right],
\label{eq:onsager}
\eea

\noindent with $A_{XY}(F) \equiv 
\partial_\alpha \widehat{R}_X \partial_\beta \widehat{R}_Y + 
\partial_\alpha \widehat{R}_Y \partial_\beta \widehat{R}_X $.
Eq.~(\ref{eq:onsager}) is evidently symmetric under a change $\alpha
\leftrightarrow \beta $, so that finally the Onsager relations,
eq.~(\ref{eq:ons}), are verified. The quantities $A_{XY}(0)$ can be obtained 
by solving eq.~(\ref{eq:te}) at stationarity for a finite number of states. 

\subsection*{a. 2--models}

For the 2--models defined in section~\ref{sec:mpaes},
the coefficient $\lambda_{12}$ can be easily calculated; it is obtained from
eq.~(\ref{eq:onsager}), by making explicit the two contributions:

\bea
& & \lambda_{12}^{(1)} = \nonumber \\ 
& & = \sum\nolimits_{XY} \frac{\chi_{_X} + \chi_{_Y}}{4}
\left( x_t - y_t \right) \left( x_c - y_c \right)
L_{XY} (0) \widehat{R}_Y (0) \nonumber \\
& & \\ 
& & \lambda_{12}^{(2)} = \sum\nolimits_{XY} \frac{\chi_{_X} + \chi_{_Y}}{4}
\frac{L_{XY}(0)}{\widehat{R}_X(0)} A_{XY}(0),
\eea

\noindent where the subscript $t$ ($c$) means translational (chemical).
After some manipulations, and bearing in mind that $x_t - y_t$ represents the
spatial displacement from state $Y$ to state $X$, while $x_c - y_c$ is the ATP
consumption from state $Y$ to state $X$ (positive when ATP is consumed), we 
obtain:

\bea
\lambda_{12}^{(1)} & = & \frac{w_1}{2} \frac{u_1+w_2}{u_1+u_2+w_1+w_2} \\
\lambda_{12}^{(2)} & = & \frac{u_1(u_2 w_2 - u_1
w_1)}{2(u_1+w_2)(u_1+u_2+w_1+w_2)}, 
\eea

\noindent where all $u_i$ and $v_i$ are calculated at $(f,\Delta \mu) = (0,0)$ 
and finally:

\be
\lambda_{12} = \frac{w_2^2 w_1 + 2u_1 w_1 w_2 +u_1 u_2
w_2}{2(u_1+w_2)(u_1+u_2+w_1+w_2)}.
\ee

Using eqs.~(\ref{eq:tr1}) and (\ref{eq:tr2}), we find:

\be
\lambda_{12} = \frac{\omega}{e^{-v}+e^{v}+2}.
\ee 

The same Onsager coefficient may be obtained by linearizing the expression for
velocity, eq.~(\ref{eq:velex}).

\subsection*{b. 3--models}

The calculation for the Onsager coefficient in 3--models by a direct application
of eq.~(\ref{eq:onsager}) is rather long. Its value may be obtained by a 
direct linearization of eq.~(\ref{eq:velex}):

\be
\lambda_{12}= \frac{\omega}{e^{-v}+2 e^{-v/2}+2 e^{v/2}+e^v+3}.
\ee

%
%
%

\end{appendix}


\end{document}